\documentclass[aps,prl,twocolumn,groupedaddress,showpacs]{revtex4}

\usepackage{graphicx}
\usepackage{dcolumn}
\usepackage{bm}
\begin{document}
\def \lafepo{LaFePO}
\def \Ba122{BaFe$_2$As$_2$}
\def \CoBa122{Ba(Fe$_{1-x}$Co$_x$)$_2$As$_2$}
\def \KBa122{Ba$_{1-x}$K$_x$Fe$_2$As$_2$}
\def \Tc{T$_c$}
\def \Ta{T$_\alpha$}
\def \Tb{T$_\beta$}

\title{Determination of the phase diagram of the electron doped superconductor Ba(Fe$_{1-x}$Co$_x$)$_2$As$_2$}


\author{Jiun-Haw Chu, James G. Analytis, Chris Kucharczyk, Ian R. Fisher}
\affiliation{Geballe Laboratory for Advanced Materials and Department of Applied Physics, Stanford University}


\date{\today}

\begin{abstract}
Systematic measurements of the resistivity, heat capacity, susceptibility and Hall coefficient are presented for single crystal samples of the electron-doped superconductor Ba(Fe$_{1-x}$Co$_x$)$_2$As$_2$. These data delineate an $x-T$ phase diagram in which the single magnetic/structural phase transition that is observed for undoped BaFe$_2$As$_2$ at 134 K apparently splits into two distinct phase transitions, both of which are rapidly suppressed with increasing Co concentration. Superconductivity emerges for Co concentrations above $x \sim 0.025$, and appears to coexist with the broken symmetry state for an appreciable range of doping, up to $x \sim 0.06$. The optimal superconducting transition temperature appears to coincide with the Co concentration at which the magnetic/structural phase transitions are totally suppressed, at least within the resolution provided by the finite step size between crystals prepared with different doping levels. Superconductivity is observed for a further range of Co concentrations, before being completely suppressed for $x \sim 0.018$ and above. The form of this $x-T$ phase diagram is suggestive of an association between superconductivity and a quantum critical point arising from suppression of the magnetic and/or structural phase transitions. 
\end{abstract}


\maketitle


\section{Introduction}

\Ba122 is a prototypical member of the new family of Fe-pnictides that play host to high temperature superconductivity. The stoichiometric compound suffers a coupled structural and antiferromagnetic transition at $\sim$ 140 K \cite{rotter}. Suppression of the broken-symmetry state by either applied pressure \cite{alireza} or chemical doping \cite{rotter-2,mandrus}  results in superconductivity, and one of the key questions associated with this entire class of material is the role played by spin fuctuations associated with the incipient tendency towards magnetism \cite{Norman, Mazin}. 

Initial work on chemical substitution in \Ba122 focused on the hole-doped material \KBa122 \cite{rotter-2}. In this case, it has been established that the structural/magnetic transition is totally suppressed for K concentrations above $x = 0.4$ \cite{xhchen-2}, whereas superconductivity appears for a wide range of concentrations from $x$ = 0.3 to 1 \cite{xhchen-2}. More recently, it has been shown that Co-substitution (i.e. \CoBa122) also suppresses the magnetic and structural transitions, eventually leading to superconductivity \cite{mandrus}. Co-doped \Ba122 is especially attractive for two reasons. First, from a fundamental point of view, this material allows an exploration of the interrelation between the structural/magnetic phase transitions with superconductivity in the context of an electron-doped system. Secondly, from a more practical perspective, crystal growth using cobalt does not suffer from the inherent difficulties associated with using potassium. Cobalt has neither a large vapor pressure nor attacks the quartz tubing used to encapsulate the growths. It is thus possible to obtain more homogeneous crystals and more reproducible synthesis conditions than is otherwise feasible for potassium-doped samples. 

In this paper we present results of resistivity, Hall coefficient, heat capacity and susceptibility measurements of single crystal samples of \CoBa122. We find that the structural/magnetic phase transition which occurs at 134 K in \Ba122 is rapidly suppressed, and splits in to two successive phase transitions with increasing Co concentration. The signatures of these two distinct phase transitions are relatively sharp, and it is unlikely that this effect is associated with phase separation due to a variation in Co concentration across an individual sample.  Superconductivity appears for Co concentrations $0.025 < x < 0.18$, with a maximum $T_c$ for a Co concentration of $x \sim 0.06$, coincident with the concentration at which the structural/magnetic phase transitions are suppressed below \Tc. The apparent coexistence of superconductivity with the broken-symmetry state on the underdoped side of the phase diagram is unlikely to be due to phase separation due to Co inhomogeneity, and mirrors what is seen for \KBa122 \cite{xhchen-2}. These measurements establish an essential symmetry between electron and hole doping of \Ba122, in which the appearance of superconductivity in both cases is clearly associated with suppression of the magnetic/structural phase transitions. Nevertheless, there are some clear differences that emerge from this study. Specifically, Co-substitution is found to be significantly more effective at suppressing the structural/magnetic phase transitions than K-substitution, and the superconducting ``dome'' extends over a much smaller range of doping. The resulting phase diagram is highly suggestive that superconductivity is intimately related to the presence of a quantum critical point associated with suppression the structural and/or magnetic phase transitions.  

\section{Results and Discussion}

Single crystals of \CoBa122 were grown from a self flux using similar conditions to published methods \cite{xhchen,mandrus}. Ba and FeAs in the molar ration 1:4 with additional Co were placed in an alumina crucible and sealed in evacuated quartz tubes. The mixture was heated to 1150C, and held for 24 hours, before slowly cooling to 1000 C, at which temperature the remaining flux was decanted using a centrifuge. The crystals have a plate-like morphology, with the $c$-axis perpendicular to the plane of the plates, and grow up to several millimeters on a side. The Co concentration was measured by electron microprobe analysis (EMPA) using undoped \Ba122 and elemental Co as standards. Measurements were made at several locations on each sample. Measured values were close to the nominal melt composition in all cases, and the variation in Co concentration across individual samples was typically characterized by a standard deviation of $0.15\%$. For instance, the Co concentration corresponding to optimal doping is $x$ = 0.061 $\pm$ 0.002. 

\begin{figure}[tbh]
\includegraphics[width=8.5cm]{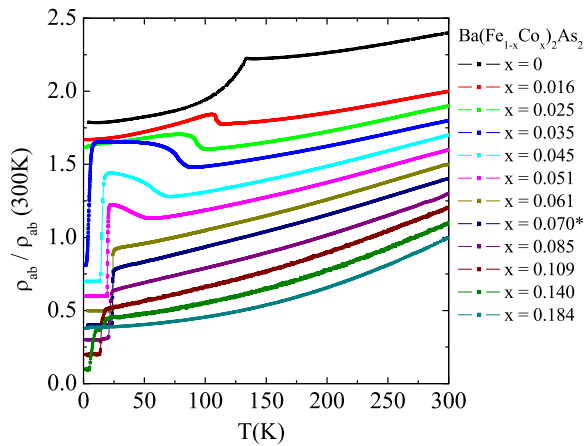}
\caption{\label{RvsT} Temperature dependence of the in-plane ($\rho_{ab}$) resistivity of \CoBa122. Data are shown normalized by the room temperature resistivity, $\rho_{ab}$(300K) , to remove uncertainty in estimates of the absolute value due to geometric factors. Successive data sets are offset vertically by 0.1 for clarity (except for $x = 0$ data, which is offset by 1.4). Values of the Co-concentration $x$ are listed in the legend, and were determined by microprobe analysis for all concentrations except the specific sample with $x = 0.070$, for which $x$ was estimated based on the observed linear relation between nominal and actual $x$.}
\end{figure}

\begin{figure}[tbh]
\includegraphics[width=8.5cm]{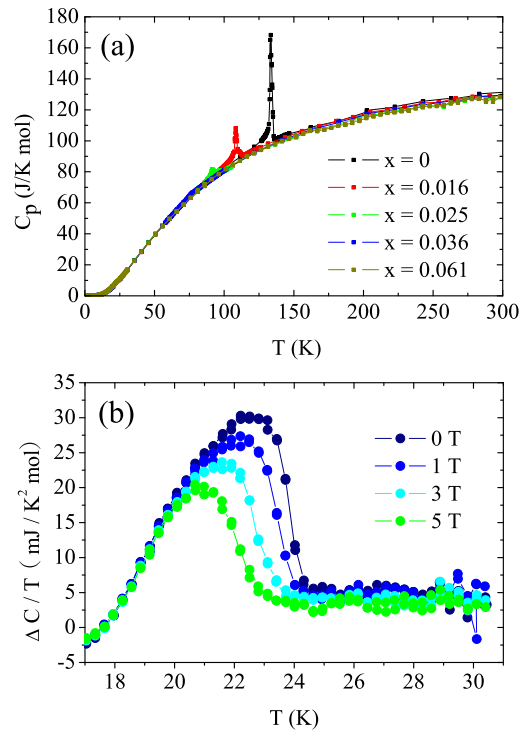}
\caption{\label{CP}(a) Heat capacity data for four representative Co concentrations ($x$ = 0, 0.016, 0.025, 0.036 and 0.061) showing the suppression of the structural/magnetic phase transitions with increasing Co concentration. (b) Superconducting anomaly for the sample with $x$ = 0.061 shown as $\Delta C/T$ vs. $T$, where $\Delta C = C_S - C_N$ and $C_N$ is estimated as described in the text from the heat capacity measured in an applied field of 14 T. Data are shown for zero field, and for applied fields of 1, 3 and 5 T along the $c$-axis.}
\end{figure}

The in-plane electrical resistivity ($\rho_{ab}$) was measured with a standard four-probe configuration using a Quantum Design Physical Properties Measurement System (PPMS), and the results are summarized in Fig. \ref{RvsT}. Data are shown normalized by the room temperature resistivity, $\rho_{ab}$(300K), to remove uncertainty in estimates of the absolute value due to geometric factors  \footnote{ Absolute values of $\rho_{ab}(300K)$ for three representative Co concentrations are 0.676 $m \Omega cm$ ($x$=0, undoped), 0.906 $m \Omega cm$ ($x$=0.061, optimally doped), and 0.696 $m\Omega cm$ ($x$=0.018, completely overdoped)}. The anomaly associated with the magnetic/structural phase transition in undoped \Ba122 is clearly visible at 134 K. This feature changes from a sharp drop in the resistivity below the transition temperature for \Ba122, to an abrupt upturn for Co-doped samples \footnote{The origin of the significant difference between the resistivity anomaly associated with the structural/magnetic transitions in undoped \Ba122 and Co-doped \Ba122 is unclear. Recent optical conductivity measurements indicate that a substantial fraction of the FS of \Ba122 is gapped below the SDW transition \cite{leo}. In this case, we can speculate that there is a subtle balance between effects associated with the reduction in the number of carriers and changes in the scattering. We note that the resistivity anomaly associated with the magnetic/structural phase transitions observed for intermediate Co concentrations is reminiscent of that which is observed for \Ba122 grown from a Sn flux \cite{canfield}, possibly indicating that Sn substitution also affects the scattering in a similarly sensitive manner. }. The anomaly is clearly suppressed in temperature with increasing $x$. For samples with $x > 0.035$ superconductivity is evident from a sharp drop to zero in the resistivity, while the anomaly associated with the structural/magnetic phase transition is still observable in the normal state. As $x$ increases beyond this value, the superconducting critical temperature \Tc\ (defined here as the midpoint of the resistive transition) increases, whereas the magnetic/structural anomaly continues to be suppressed. Finally \Tc\ reaches the highest value of 24 K in the sample with $x = 0.061$ (``optimal" doping), for which the normal state anomaly is no longer observable. Further increasing $x$ beyond this value results in an eventual suppression of superconductivity (``overdoping"). For the sample with $x = 0.184$, \Tc\ is below our instrumental base temperature of 1.8 K. 

The suppression of the magnetic/structural phase transitions was also followed by heat capacity, measured for single crystal samples by a relaxation technique using a Quantum Design PPMS. Results are shown in Fig. \ref{CP}(a) for four representative Co concentrations. For undoped \Ba122, a sharp peak at 134 K marks the first order structural/magnetic phase transition as reported previously \cite{dong}. With increasing $x$ both the transition temperature and the magnitude of the peak are rapidly suppressed, until the anomaly becomes almost unobservable for $x > 0.036$. In order to provide an estimate of the phonon background, the heat capacity was also measured for an optimally doped sample (i.e. $x$ = 0.061, which is the smallest Co concentration for which the magnetic/structural phase transition is totally suppressed). These data were taken to lower temperature, revealing the superconducting anomaly. Data for this particular Co-concentration are shown in Fig. \ref{CP}(b) as $\Delta C/T$ vs. $T$, where $\Delta C$ is the difference between the heat capacity in the superconducting state $C_S$ and the normal state $C_N$. The latter quantity has been estimated over the plotted temperature range from measurements made in a field of 14 T applied along the $c$-axis. These data indicate $\Delta C/T_c$ $\approx$ 23 (mJ/K$^2$ mol) in zero field.

\begin{figure}[tbh]
\includegraphics[width=8.5cm]{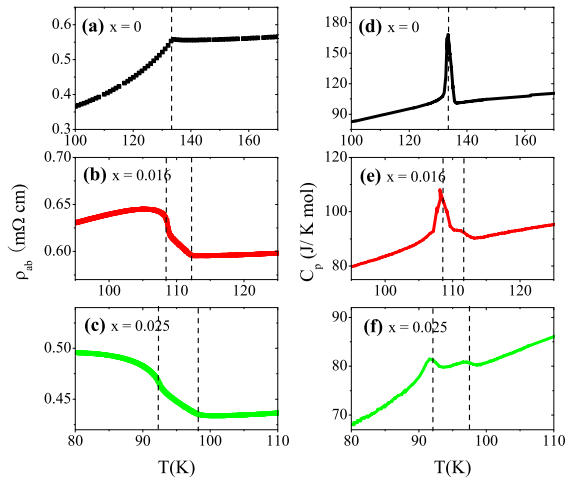}
\caption{\label{zoom}Resistivity (panels a-c) and heat capacity (panels d-f) data for samples with $x$ = 0, 0.016 and 0.025 over a narrower temperature interval, showing the splitting of the single phase transition observed for $x$ = 0 in to two distinct phase transitions for $x > 0$.}
\end{figure}

Careful inspection of the resistivity and heat capacity data shown in Fig.s \ref{RvsT} and \ref{CP} reveal that the single anomaly associated with the magnetic/structural phase transition which is seen for undoped \Ba122 splits in to two distinct features for the Co-doped samples. In fig \ref{zoom} we show the resistivity and heat capacity anomaly near these transitions for the undoped and two lightly doped samples. It is evident that the single sharp downturn in \Ba122 becomes two successive sharp changes in slope in the resistivity of the lightly doped \CoBa122.  As for the heat capacity, the single sharp first-order peak observed for undoped \Ba122, becomes two distinct features, resembling a second order-like step followed by a lower temperature peak, reminiscent of a broadened first-order transition.

\begin{figure*}[tbh]
\includegraphics[width=17cm]{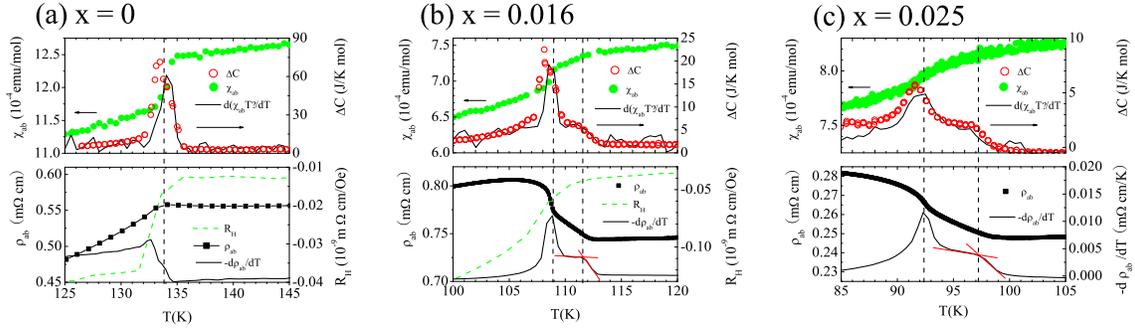}
\caption{\label{G3} Comparison of heat capacity, susceptibility, resistivity and Hall coefficient data for samples with (a) $x = 0$, (b) $x = 0.016$ and (c) $x = 0.025$. In the upper panels, the heat capacity (right axis) is shown as $\Delta C$, having subtracted the background phonon contribution as described in the main text, together with the in-plane susceptibility $\chi_{ab}$ (left axis) and the derivative $d(\chi T)/dT$ (arb. units). In the lower panels, absolute values of the in-plane resistivity $\rho_{ab}$ (left axis) are shown together with the Hall coefficient $R_H$ (right axis) and the derivative -$d\rho/dT$ (arb units). Vertical dotted lines mark successive phase transitions, evident in all four physical properties.}
\end{figure*}

\begin{figure}[tbh]
\includegraphics[width=8.5cm]{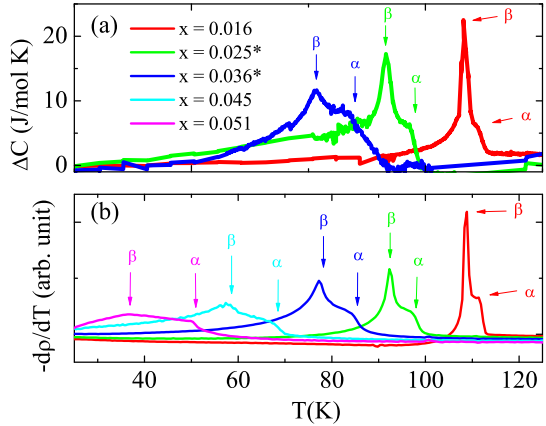}
\caption{\label{CPDR}Progression of the two successive phase transitions, marked as $\alpha$ and $\beta$, as a function of Co concentration as observed in (a) heat capacity and (b) the derivative of the resistivity -$d\rho/dT$. Heat capacity data are shown as $\Delta C$, having subtracted the background phonon contribution as described in the main text. Data for $x$ = 0.025 and 0.036 have been multiplied by factors of 3 and 5 respectively for clarity.}
\end{figure}

To further investigate the apparent splitting of the structural/magnetic phase transitions, magnetic susceptibility and Hall coefficient data were also measured. Results are shown in Fig \ref{G3} for two representative Co concentrations. Significantly, both measurements indicate the presence of two distinct phase transitions for $x > 0$ evidenced by distinct breaks in the slope of each quantity. In the case of the susceptibility, this effect is more clearly apparent as two peaks in the derivative $d(\chi T)/dT$ \footnote{Susceptibility data were smoothed using a linear interpolation and adjacent points averaging before taking the derivative.}. These data can be compared with the heat capacity measurements (right axis of the upper panels), where we have explicitly subtracted the normal state heat capacity of the optimal doped \CoBa122 with $x = 0.061$ as described above, to provide an estimate of the contribution to the heat capacity arising from the two phase transitions. The background subtracted heat capacity is in excellent agreement with the derivative $d(\chi T)/dT$ of the susceptibility, which is proportional to the heat capacity near a second-order phase transition\cite{fisher-1962}. These data are also in excellent agreement with the derivative of the resistivity $d\rho/dT$. In short, all four measurements indicate unambiguously the presence of two distinct phase transitions. It is very unlikely that this splitting arises from Co inhomogeneity, which in it's simplest form would cause a continuous broadening rather than a distinct splitting of the two phase transitions. Indeed, the presence of Co inhomogeneity can be inferred from the broadening of each of these two phase transitions (Fig. \ref{CPDR}), but this effect remains less than the actual splitting.

Although the magnitude of the heat capacity anomalies associated with these two phase transitions is rapidly suppressed for $x > 0.036$, nevertheless the close correspondence between the heat capacity and the resistivity derivative can be used to follow these phase transitions to even higher Co concentrations. Data for Co concentrations up to $x$ = 0.051 are shown in Figure \ref{CPDR}. As can be seen, both transitions, marked as $\alpha$ and $\beta$, are broadened with increasing Co concentration, but the two specific features which can be associated with the critical temperatures for $x$ = 0.016, 0.025 and 0.036 remain visible in the resistivity derivative even for $x$ = 0.045 and 0.051. For larger values of the Co concentration, no phase transitions are evident above $T_c$. 

\begin{figure}[tbh]
\includegraphics[width=8.5cm]{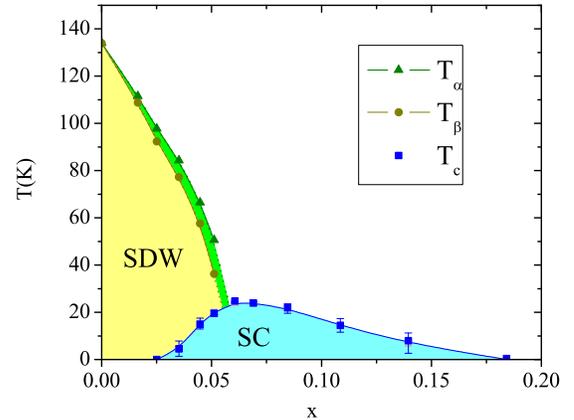}
\caption{\label{Phase}Phase diagram for \CoBa122 showing the suppression of the two successive phase transitions $\alpha$ and $\beta$ with increasing Co concentration, and the eventual superconducting (SC) "dome". Data points for $T_{\alpha}$ and $T_{\beta}$ were obtained from heat capacity, resistivity, Hall coefficient and susceptibility data for $x$ = 0, 0.016, 0.025 and 0.036, and from resistivity data alone for $x$ = 0.045 and 0.051. Superconducting $T_c$ values were obtained from resistivity data. Error bars indicate 10 and 90$\%$ of the resistive transition. Uncertainty in the Co concentration corresponds to less than the width of the data markers.}
\end{figure}

Based on the measurements described above, we can establish a composition-temperature ($x-T$) phase diagram for \CoBa122, shown in Fig. \ref{Phase}. Phase boundaries delineate the successive phase transitions $T_{\alpha}$, $T_{\beta}$ and $T_c$ as described above. In analogy to the oxy-pnictide LaFeAsO \cite{Dai-1} it is tempting to speculate that the upper transition $T_{\alpha}$ corresponds to the structural distortion, while the lower transition $T_{\beta}$ corresponds to the SDW transition, but the current data are not sufficient to make this association and further scattering and/or microscopic measurements will be necessary to determine the nature of the two transitions.

The apparent coexistence of superconductivity with the broken symmetry SDW state on the underdoped side of the phase diagram raises the question of whether there is macroscopic phase separation associated with a variation in the Co concentration. For length scales above one micron this can be ruled out based on the microprobe analysis. Specifically, the standard deviation in the cobalt doping is much smaller than the range across which we observed this coexistence. From a materials viewpoint it would also be surprising to find Co inhomogeneity on shorter length scales given the large solid solubility. This does not, however, rule out the intriguing possibility of a spontaneous phase separation in an otherwise homogeneous material. In this regard susceptibility measurements provide some insight. 

\begin{figure}[tbh]
\includegraphics[width=8.5cm]{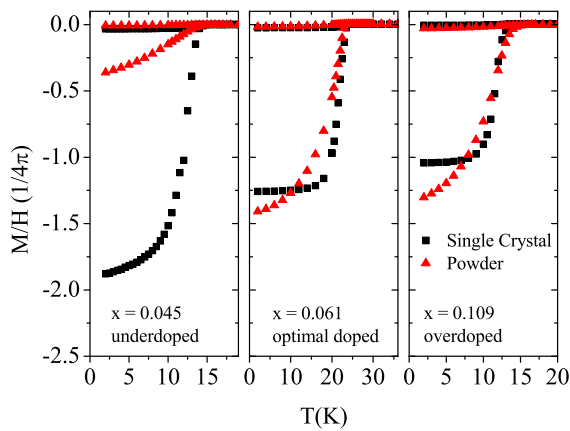}
\caption{\label{fczfc} Temperature dependece of susceptibility following zero-field cooling (zfc) and field cooling (fc) cycles, for both single crystal and powdered samples for three representative Co concentrations corresponding to $x = 0.045$ (underdoped), $x = 0.061$ (optimally doped) and $x = 0.109$ (overdoped). Measurements were made in a field of 50 Oe applied parallel to the $ab$-plane of the single crystals.}
\end{figure}

Susceptibility data were taken for representative Co concentrations on the underdoped and overdoped side of the phase diagram, as well as for an optimally doped sample. These measurements were made for applied fields of 50 Oe, oriented in the $ab$-plane, for both zero-field cooling and field cooling cycles. The measurements were then repeated for each sample after grinding to a fine powder to examine the role of screening currents. The results of these measurements are shown in Fig \ref{fczfc}. First we consider the single crystal data (black data points). Field cooled values are relatively small for each Co concentration, indicative of substantial flux trapping. However, zero-field cooled values are large for both underdoped, optimally doped and overdoped cases indicating the presence of significant screening currents. These values do not change substantially when the single crystals are ground to fine powders (red data points) for the optimally doped and overdoped cases, but significantly the underdoped material has a substantially smaller value for the zfc susceptibility. It is clear that the underdoped material is still a bulk superconductor. However, the ability of the underdoped material to support screening currents is significantly reduced by powdering. This effect might be related to the carrier concentration. Specifically, the superfluid density will presumably be smaller for the underdoped material given the coexistence with the SDW state, leading to larger values for the penetration depth. In this case, it is possible that some fraction of the grains in the finely powdered sample will have a size smaller than the penetration depth, reducing the zfc susceptibility. Equally, it is possible that powdering causes significant damage to the surface of the resulting grains. The asymmetry of the superconducting dome might then lead to a stronger suppression of superconductivity for underdoped relative to overdoped samples. Most intriguingly, the lower zfc susceptibility might also reflect the presence of intrinsic inhomogeneity on the underdoped side of the phase diagram, because the percolating shielding currents do not  
encompass the entire sample. The current measurements are not sufficient to determine the origin of this rather dramatic effect on the susceptibility. Nevertheless, it is clear that there is a substantial volume fraction even in the underdoped region of the phase diagram, but that the superconductivity is to some extent more fragile, at least to the effect of powdering. 

The phase diagram shown in Fig. \ref{Phase} is highly suggestive of a scenario in which the superconductivity is intimately related to the presence of a quantum critical point associated with the eventual suppression of the structural and/or magnetic phase transitions. Clearly this needs to be investigated in greater detail. Nevertheless, this observation raises the question of whether a similar mechanism is at work in the hole doped analog \KBa122, and more broadly for all of the superconducting Fe-pnictides. In the specific case of \Ba122, Co-doping is clearly more effective at suppressing the magnetic/structural transitions than K-doping, but it remains to be determined whether this is related to an asymmetry in the electronic structure or to other possibilities.

During preparation of this manuscript we became aware of two similar studies of the effects of Co-substitution in \Ba122 recently posted on the ArXiv.org preprint server. In the first of these, Ni \textit{et. al.} map out a similar phase diagram, also finding a splitting of the magnetic/structural phase transition \cite{canfield-2}. These results broadly corroborate our findings. In the second, Ning \textit{et. al.} use NMR measurements to follow spin dynamics for representative underdoped and overdoped samples, also surmising the presence of a quantum critical point \cite{mandrus-2}.

\section{Conclusion}
In conclusion, we have determined the phase diagram for the electron-doped superconductor \CoBa122. We find that the single structural/magnetic phase transition that occurs in \Ba122 splits with Co-doping, although the nature of the two split transitions remains to be determined. Critical temperatures associated with both transitions are progressively reduced with increasing Co concentration, and are suppressed below $T_c$ coincident with optimal doping. The superconducting "dome" extends over a limited range of Co concentrations, from $x$ = 0.025 to $x$ = 0.18. These observations clearly delineate regions of the $x-T$ phase diagram for which further experiments have the potential to reveal the interplay between the structural/magnetic and superconducting phase transitions.   

\section{Acknowledgements}
The authors thank R. E. Jones for technical assistance with EMPA measurements, and T. H. Geballe, M. R. Beasley and L. DeGiorgi for helpful discussions. This work is supported by the DOE, Office of Basic Energy Sciences, under contract no. DE-AC02-76SF00515.

\end{document}